\documentclass[10pt]{article}
\usepackage{fullpage}
\usepackage{amsmath}
\usepackage{amssymb}
\usepackage{amsfonts}
\usepackage[dvips]{epsfig}
\usepackage{color}
\usepackage{cite}
\graphicspath{{figures/}}

\begin{document}

\begin{center}

\LARGE{ {\bf Statistical distribution of quantum correlation induced by multiple scattering in the disordered medium
}}
\end{center}
\begin{center}
\vspace{10mm} \large

{\bf Dong Li}$^{1,2,}$\footnote{Corresponding author. E--mail: lidong@mtrc.ac.cn}, {\bf Yao Yao}$^{1,2,}$\footnote{Corresponding author. E--mail: yaoyao@mtrc.ac.cn} and
{\bf Mo Li}$^{1,2}$ 

\vspace{2mm}

$^1$\emph{Microsystems and Terahertz Research Center, China Academy of Engineering Physics, Chengdu Sichuan 610200, China}

 \vspace{2mm}

 $^2$\emph{Institute of Electronic Engineering, China Academy of Engineering Physics, Mianyang Sichuan 621999, China}

\vspace{5mm}
\normalsize

\end{center}

\begin{center}
\vspace{15mm} {\bf Abstract}
\end{center}

For the quantum correlations between scattered modes in the disordered media, the previous works focus mainly on the cases where the inputs are non-superposed states, for instance, products of Fock states [Phys. Rev. Letts. 105 (2010) 090501]. A natural question that arises is how the superpositions affect the quantum correlations. Following this trail, the comparison between superpositions and products of Fock states is performed. It is found an interesting phenomenon that for the superposition and the corresponding product of Fock state (non-Gaussian states), their averaged quantum correlations are nearly same, whereas the distributions of their quantum correlations might be different. Therefore, superpositions may affect the distributions of the quantum correlations. In addition, to examine how the Gaussian states affect the quantum correlations, we compare the typical Gaussian states with the non-Gaussian states (superpositions and products of Fock states). It is discovered that the non-Gaussian-state input could result in the quantum correlation that is either positive or negative, depending on the number of the input modes and the number of the photons in each mode, whereas the Gaussian-state input always leads to the non-negative quantum correlation. Besides, it is demonstrated that with the increase of the disorder strength, the mean strength of the quantum correlation increases for multi-mode-state inputs (except for multi-mode-coherent-state inputs). These results may be useful to control and adjust the quantum properties of scattered modes after the quantized lights propagating through the disordered medium.

\vspace{5mm}

\vspace{5mm}

\section{Introduction}

The studies of the multiple scattering for quantized lights in a disordered medium have received long-sustained attentions \cite{b1998,patra2000,two2002,wiersma2013,lahini2010,defienne2016,leonetti2013,starshynov2016,an2018,walschaers2016,zhang2018}, which leads to a bunch of novel phenomena in the quantum optics theory, such as, the quantum noise transmitted through a multiple scattering medium \cite{lodahl2005a,scalia2013}, the identification of frequency \cite{lodahl2006a} and spatial \cite{lodahl2005b,gilead2015,schlawin2012} quantum correlations (QCs), among the others. Besides, this quantum optical system, nonclassical lights illuminating on a disordered medium, has a great potential for numerous applications in quantum information processing, including programmable quantum optical circuit \cite{wolterink2016,huisman2014,defienne2014}, Heisenberg-limit resolution imaging \cite{hong2017,hong2018}, quantum optical authentication \cite{goorden2014,nikolopoulos2017,yao2016,li2017}, and quantum communication \cite{b2017}. In particular, the QC between the scattered modes, as a measure of quantum interference between input modes \cite{ott2010}, is usually regarded as a significant index to characterize the quantum optical properties of a disordered medium.

Initially, the pioneering works established the related theoretical frame based on the ensemble-averaged QC over all realizations of disorder. For instance, in 2005, a new spatial QC was theoretically predicted in light transport through a multiple scattering medium where the QC depends on the quantum input state of the light \cite{lodahl2005b}. In 2006, the QC was theoretically investigated in the disordered medium with the quadrature squeezed light as input \cite{lodahl2006b}. It was shown that the QC appeared in the number of photons but not in the quadrature amplitudes. Later, in 2010, Ott \textit{et al.} \cite{ott2010} reported on the effects of quantum interference induced by transmission of an arbitrary number of products of Fock states through a disordered medium. More recently, in 2016, Starshynov \textit{et al.} \cite{starshynov2016} studied theoretically how multiple scattering of Gaussian states in a disordered medium can spontaneously generate the QCs. 

Meanwhile, the significant progress was also made in experiments. For example, Smolka \textit{et al.} \cite{smolka2009} observed the spatial QC between the scattered modes in the disordered medium experimentally in 2009, where the mean strength of the quantum correlation is related to the number of incident photons. In 2010, Peeters \textit{et al.} \cite{peeters2010} reported the observation of the QC in the speckle patterns with spatially entangled photon pairs as inputs. They identified that the two-photon speckle has a much richer pattern than the ordinary one-photon speckle.

Despite these important breakthroughs, the previous literatures concentrate mainly on the non-superposed states, such as, the products of Fock states \cite{ott2010}. A natural question that arises is how the superposition affects the quantum correlation. Following this line, we compare the two cases: the superpositions and products of Fock states. It is discovered that the superpositions and the corresponding products of Fock states have nearly same averaged quantum correlation while they may present different distributions of quantum correlations. Therefore, the superpositions may affect the distributions of quantum correlations. 

Nevertheless, both the products and superpositions of Fock states are non-Gaussian states. For the completeness of this manuscript, we wonder how the Gaussian states affect the quantum correlations. Thus the comparison between typical Gaussian states (coherent, thermal, and squeezed-vacuum states) and non-Gaussian states (products or superpositions of Fock states) is performed. It is demonstrated that the non-Gaussian-state input could result in the quantum correlation that is either positive or negative, depending on the number of the input modes and the number of the photons in each mode, whereas the Gaussian-state input always leads to the non-negative quantum correlation. Besides, it is discovered that with the increase of the disorder strength, the mean strength of the quantum correlation increases for the multi-mode-state inputs (except for the multi-mode-coherent-state inputs).

This manuscript is organized as follows: Sec. 2 briefly describes the model for the propagation of the quantized lights through a disordered medium. Then, in Sec. 3, it presents the distribution of the QC induced by multiple scattering in different-disordered media with products and superpositions of Fock states as inputs. Sec. 4 compares the QCs between the Gaussian and non-Gaussian input states. Finally, Sec. 5 is devoted to the conclusion of the main results.

\section{Theoretical model}
\label{sec2}

\subsection{Propagation of the quantized light through a disordered medium}

\begin{figure}[htbp]
\begin{center}
\includegraphics[width=.50\textwidth]{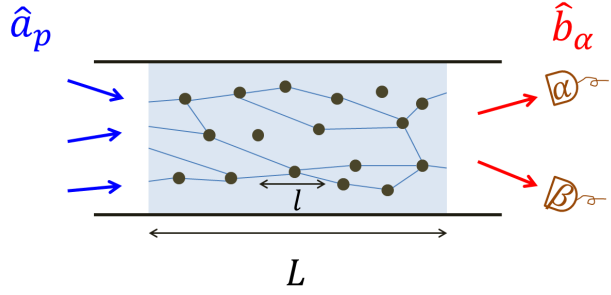} {}
\end{center}
\caption{Transport of the quantized light through a disordered medium with the length of $L$ and the mean free path of $l$. $\hat{a}_{p}$ denotes the input mode and $\hat{b}_{\alpha}$ the output mode. Two arbitrary output spatial modes ($\alpha$ and $\beta$) are monitored by two detectors}
\label{fig1setup}
\end{figure}

Fig. \ref{fig1setup} describes the propagation of the quantized light through a disordered medium with the length of $L$, the mean free path of $l$ and the number of transmission channels of $N$. The blue (red) arrows represent the input (output) modes and $\hat{a}_{p}$ $(\hat{b}_{\alpha})$ denotes the photon annihilation operator of input (output) modes. We define $s \equiv L/l$ which determines the degree of disorder.

The random-matrix theory of quantum transport \cite{beenakker1997,rossum1999,garcia2002,cwilich2006,rotter2017,dorokhov1982,xu2017a,xu2017b} is adopted to simulate the propagation behavior, in which the scattering matrix denoted by $S$ describes clearly the input-output relation. After transport through a disordered medium, related to the initial modes, the final outgoing modes can be expressed as:
\begin{equation}
\hat{b}_{\alpha} = \sum_{p}S_{\alpha p} \hat{a}_{p}, 
\label{scatterequ}
\end{equation}
where $S_{\alpha p}$ is the element of the $S$ matrix which represents the transmission coefficient from the incoming modes to the outgoing modes. Obviously, it is found that each spatial output mode is dependent on the all incident modes according to Eq. (\ref{scatterequ}). For brevity, the method to generate random scattering matrix $S$ is shown in Append. \ref{rmtappd} in detail. 

\subsection{Quantum correlation between the scattered modes}

The QC between two arbitrary output spatial modes $\alpha$ and $\beta$ is related to the probability of detecting the arrival of a photon at mode $\alpha$ and another photon at mode $\beta$ \cite{walls2007}. In experiments, the corresponding QC can be measured by coincidence counting detection on modes $\alpha$ and $\beta$ as depicted in Fig. \ref{fig1setup}. In theory, the QC \cite{ott2010} can be expressed as
\begin{equation}
C_{\alpha \beta} = \frac{\langle:\hat{n}_{\alpha} \hat{n}_{\beta}:\rangle}{\langle \hat{n}_{\alpha}\rangle \langle \hat{n}_{\beta} \rangle} - 1, 
\label{qcorr}
\end{equation}
where $\hat{n}_{\alpha} = \hat{b}^{\dagger}_{\alpha} \hat{b}_{\alpha} $ and $\hat{n}_{\beta} = \hat{b}^{\dagger}_{\beta} \hat{b}_{\beta} $ are the photon number operators of output modes $\alpha$ and $\beta$, respectively, $::$ represents the normal ordering, and the brackets denote quantum mechanical expectation values [$\hat{b}_{k}^{\dagger}$ ($\hat{b}_{k}$) denoting the creation (annihilation) operator of scattered mode $k$ ($k = \alpha, \beta$)]. It is worthy pointing out that $C_{\alpha \beta} < 0$ $(C_{\alpha \beta} > 0$ or $C_{\alpha \beta}=0)$ represents the cases where the output modes $\alpha$ and $\beta$ are anticorrelated (correlated or uncorrelated). 

To present a full picture of the QC, we analyze the statistical distribution, $P(C_{\alpha \beta})$, of the QC. In order to determine the distribution $P(C_{\alpha \beta})$ accurately, a large number of samples of random scattering matrices is considered where each random scattering matrix corresponds to a realization of disorder. Each realization provides a specific set of $C_{\alpha \beta} $ from which we can calculate the distribution $P(C_{\alpha \beta})$. Moreover, to characterize the distribution of the QC, it is convenient to introduce the ``deviation" of the QC 
\begin{equation}
{{\Delta}}C_{\alpha \beta} = {{\Delta}}C_{\alpha \beta}^P + {{\Delta}}C_{\alpha \beta}^N,
\end{equation}
where the ``deviation" is composed of two parts: the positive ``deviation" [${{\Delta}}C_{\alpha \beta}^P \equiv \sqrt{\overline{(C_{\alpha \beta} - \bar{C}_{\alpha \beta})^2}}$ if $C_{\alpha \beta} > \bar{C}_{\alpha \beta}$] and the negative ``deviation" [${{\Delta}}C_{\alpha \beta}^N\equiv \sqrt{\overline{(C_{\alpha \beta} - \bar{C}_{\alpha \beta})^2}}$ if $C_{\alpha \beta} < \bar{C}_{\alpha \beta}$] with $\bar{j}$ denoting the average of $j$.

\section{Numerical results}

Inserting Eq. (\ref{scatterequ}) into Eq. (\ref{qcorr}), one can rewrite the QC as
\begin{equation}
C_{\alpha \beta} = \frac{\sum_{mnpq} S_{\alpha m}^{\ast} S_{\beta p}^{\ast} S_{\beta q} S_{\alpha n} \langle \hat{a}_m^{\dagger} \hat{a}_{p}^{\dagger} \hat{a}_{q} \hat{a}_n \rangle}{\left(\sum_{mn} S_{\alpha m}^{\ast} S_{\alpha n} \langle \hat{a}^{\dagger}_{m} \hat{a}_{n} \rangle\right) \left( \sum_{pq} S_{\beta p}^{\ast} S_{\beta q} \langle \hat{a}^{\dagger}_{p} \hat{a}_{q} \rangle \right)}-1,
\label{cab}
\end{equation}
where $S_{ij}^{\ast}$ ($i = \alpha, \beta$ and $j = m, n, p, q$) denotes the conjugate of $S_{ij}$ and $\hat{a}_j^{\dagger}$ ($\hat{a}_j$) indicates the creation (annihilation) operator of input mode $j$. According to Eq. (\ref{cab}), one can easily evaluate the corresponding QC when given the input state and the scattering matrix.

\subsection{Products of Fock states as inputs}

{\bf Single-mode Fock state}  ${|n\rangle}$: If the input is in a single-mode Fock state $|n\rangle$ \cite{noteinput}, Eq. (\ref{cab}) always gives a constant value of $C_{\alpha \beta} = - 1/\bar{n}$ that is independent of the degrees of disorder $s$, where $\bar{n}=n$ is the mean photon number. In particular, when $\bar{n}=2$, $C_{\alpha \beta} = -1/2$ neatly reproduces the results in Ref. \cite{ott2010}. If $\bar{n}$ is finite, $C_{\alpha \beta} = -1/\bar{n} < 0$ indicates that the scattered modes are anticorrelated, and it becomes completely uncorrelated at the limit of $\bar{n} \rightarrow +\infty$.

\begin{figure}[tbhp]
\begin{center}
\includegraphics[width=.680\textwidth]{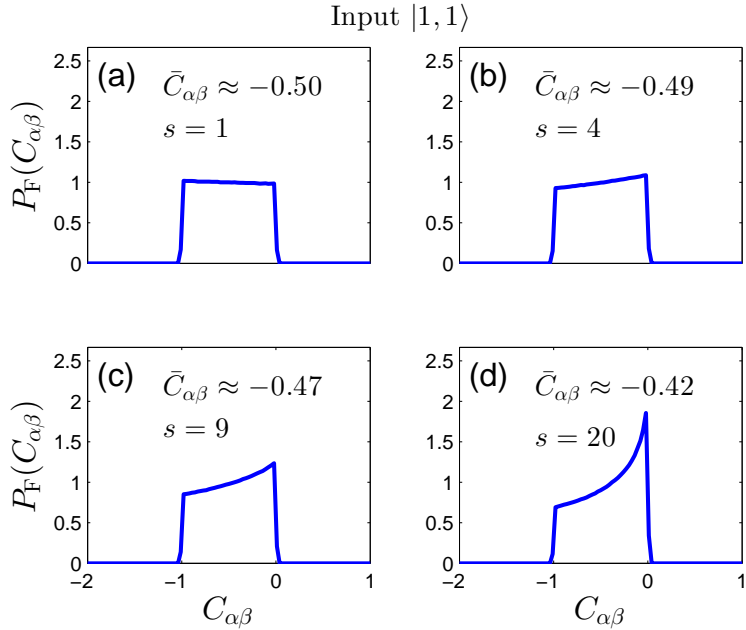} {}
\end{center}
\caption{The normalized probability distribution of the QC with the input of $|1,1\rangle$ for different disorder strengths: (a) $s = 1$, (b) $s = 4$, (c) $s = 9$, and (d) $s = 20$. In order to obtain an accurate statistics, the results presented here involves $10^7$ samples of generated scattering matrices with respect to each $s$. Parameters used are: the number of transmission channel $N=20$ and $\int P_{\rm{F}}(C_{\alpha \beta}) d C_{\alpha \beta} = 1$}
\label{fig11two}
\end{figure}

{\bf Two-mode-single-photon Fock state} ${|1,1\rangle}$: Fig. \ref{fig11two} shows the normalized statistical probability distribution of the QC, $P_{\rm{F}}(C_{\alpha \beta})$, with the state of $|1,1\rangle$ as input. To investigate the effects of the disorder on the QC, four different values of $s$ ($s=1,4,9$, and $20$) have been analyzed as depicted in Fig. \ref{fig11two}. It can been seen that $C_{\alpha \beta}$ is dependent on the degree of disorder $s$ which is different from the case of the single-mode Fock states independent on $s$. Moreover, as shown in Figs. \ref{fig11two}(a)-(d), with $s$ increases, the anticorrelation gradually becomes weak since $C_{\alpha \beta}$ tends towards zero. This indicates that for the random media of large degree of disorder, the probability of two photons arriving at two different output modes $\alpha$ and $\beta$ increases, although they are still anticorrelated ($C_{\alpha \beta}<0$).

\begin{figure}[tbhp]
\begin{center}
\includegraphics[width=.680\textwidth]{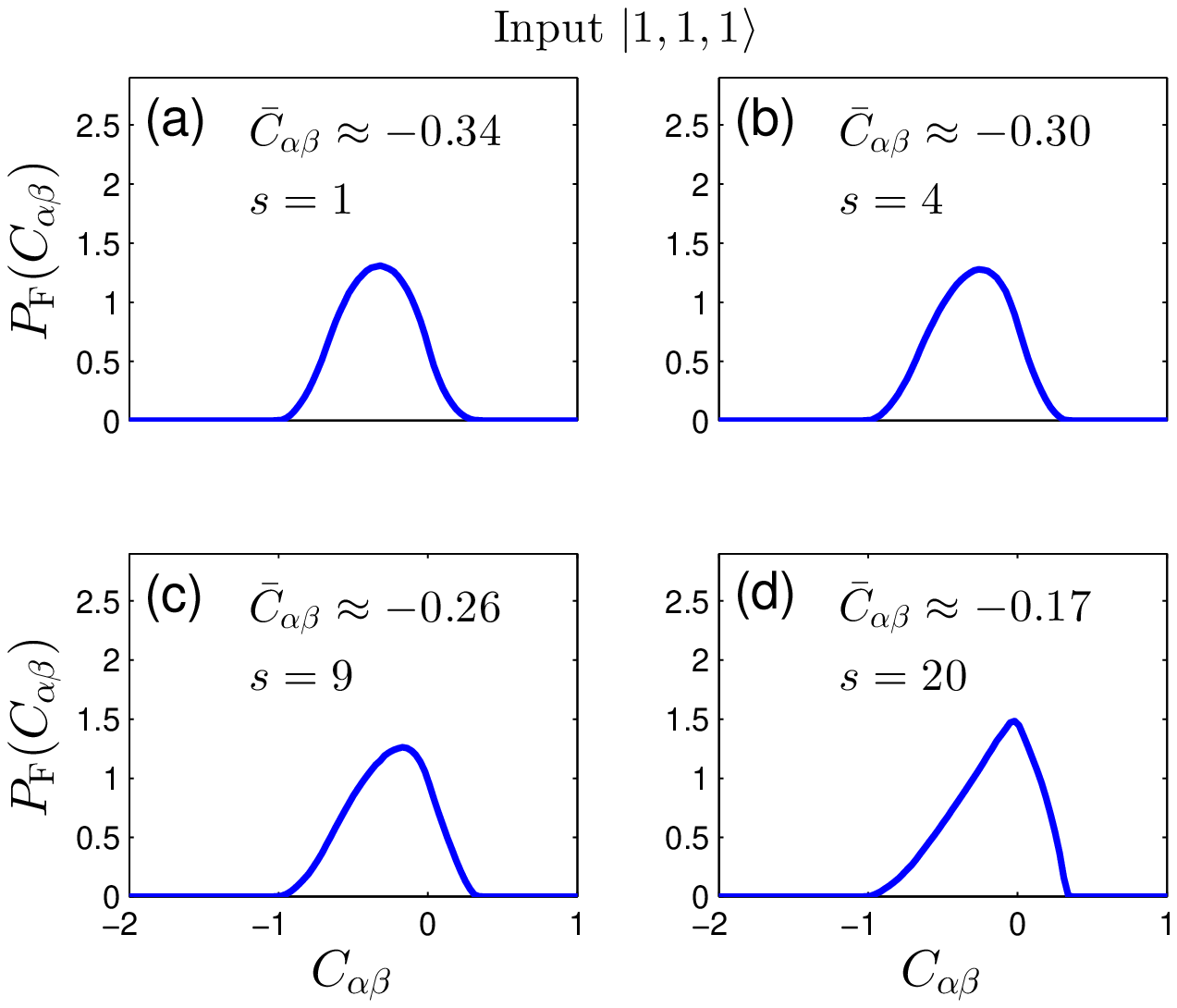} {}
\end{center}
\caption{The normalized probability distribution of the QC with the input of $|1,1,1\rangle$ for different disorder strengths: (a) $s = 1$, (b) $s = 4$, (c) $s = 9$, and (d) $s = 20$. All the other parameters are the same as the case in Fig. \ref{fig11two}}
\label{fig111three}
\end{figure}

{\bf Three-mode-single-photon Fock state} ${|1,1,1\rangle}$: When the number of the input modes increases as $|1,1,1\rangle$, it is possible to obtain a positive QC ranging between $-1$ and around $0.4$ for $s=1,4,9$, and $20$ as shown in Fig. \ref{fig111three}. This result indicates that the output modes can be either anticorrelated ($C_{\alpha \beta} < 0$) or correlated ($C_{\alpha \beta} >0$). Comparing with the two-mode case $|1,1\rangle$, it is clear that the number of modes plays a vital role in the distribution of the QC, Interestingly, as the disorder $s$ increases, the central peak of $C_{\alpha \beta}$ moves to the right, showing that the output modes are more likely to be correlated in the large-disordered media, which is consistent with the two-mode results.

\subsection{Superpositions of Fock states as inputs}

It has been investigated that the case of products of Fock states is considered as input whereas it is still missing for superpositions of Fock states. Naturally, a new question arises from how the superpositions affect the QC, since superposition is considered as a vital quantum resource which lies at the heart of numerous nonclassical properties of quantum mechanics \cite{yao2015,superposition}.

In order to compare with the product of $|1,1\rangle$ ($|1,1,1\rangle$), one considers the superposition of $|\rm{SP}2\rangle$ ($|\rm{SP}3\rangle$) as inputs where $|\rm{SP}2\rangle \equiv (|2,0\rangle + |0,2\rangle) / \sqrt{2}$ ($|\rm{SP}3\rangle \equiv (|3,0,0\rangle + |0,3,0\rangle + |0,0,3\rangle) / \sqrt{3}$) and the SP denotes superposition. It is worthy to note that the state of $|\rm{SP}2\rangle$ ($|\rm{SP}3\rangle$) has the same mean photon number as $|1,1\rangle$ ($|1,1,1\rangle$). 

\begin{figure}[tbhp]
\begin{center}
\includegraphics[width=.680\textwidth]{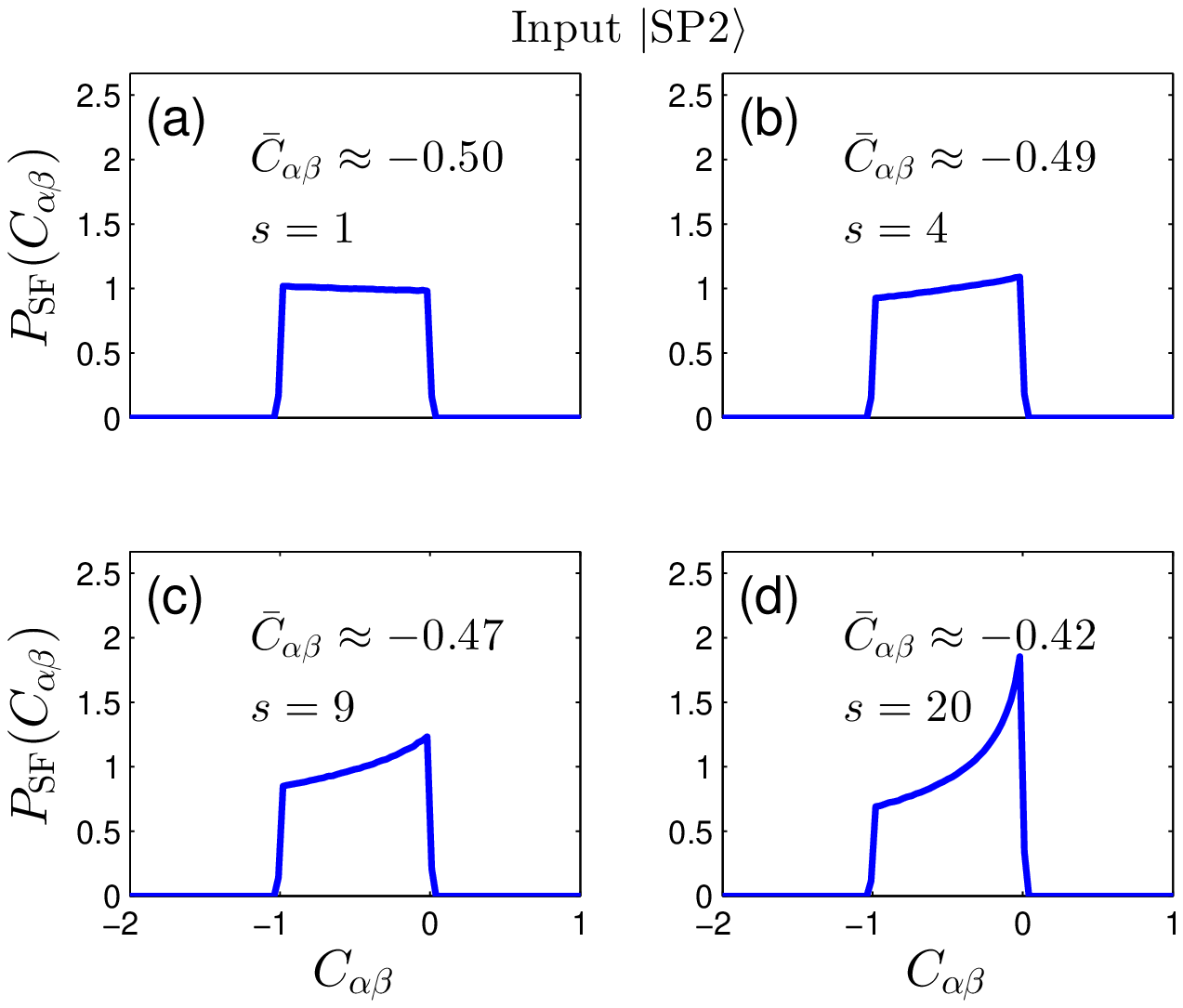} {}
\end{center}
\caption{The normalized probability distribution of the QC with the input of $|\rm{SP}2\rangle$ for different disorder strengths: (a) $s = 1$, (b) $s = 4$, (c) $s = 9$, and (d) $s = 20$. All the other parameters are the same as the case in Fig. \ref{fig11two}}
\label{fig0220}
\end{figure}

{\bf Superposition state} ${|\textbf{SP}2\rangle}$: Fig. \ref{fig0220} plots the distribution of the QC with the input of $|\rm{SP}2\rangle$. When $s=1$, $C_{\alpha \beta}$ is distributed uniformly between $-1$ and $0$, where its average $\bar{C}_{\alpha \beta} \approx 0.50$ as shown in Fig. \ref{fig0220}(a). With $s$ increasing, $C_{\alpha \beta}$ tends to be close to zero which yields that the anticorrelation fades gradually. Meanwhile, the averaged QC, $\bar{C}_{\alpha \beta}$, increases with the increase of $s$. Comparing between Figs. \ref{fig0220} and \ref{fig11two}, one can find that both distributions and averages of the QCs of $|\rm{SP}2\rangle$ are similar to the case of $|1,1\rangle$. 

\begin{figure}[tbhp]
\begin{center}
\includegraphics[width=.680\textwidth]{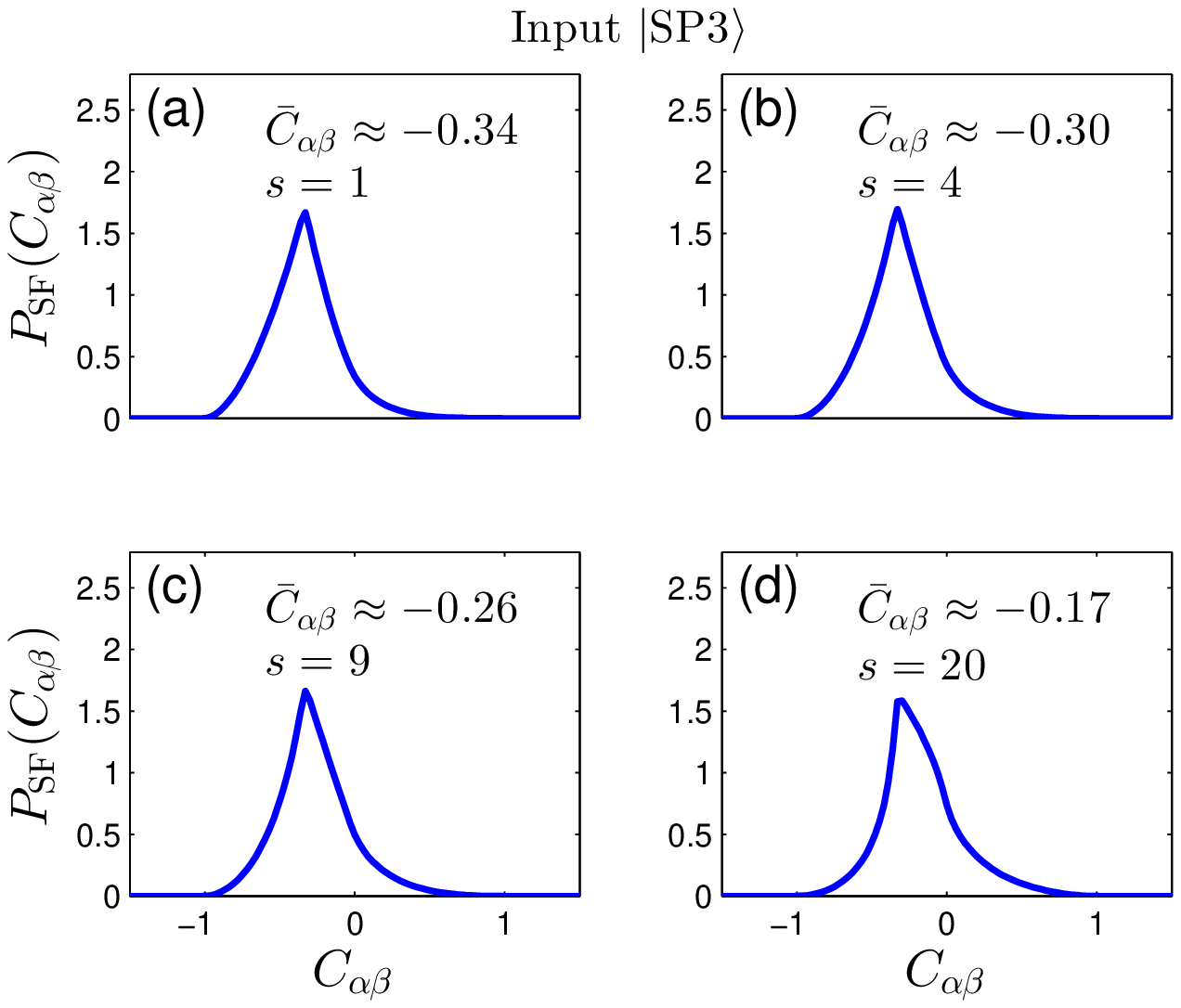} {}
\end{center}
\caption{The normalized probability distribution of the QC with the input of $|\rm{SP}3\rangle $ for different disorder strengths: (a) $s = 1$, (b) $s = 4$, (c) $s = 9$, and (d) $s = 20$. All the other parameters are the same as the case in Fig. \ref{fig11two}}
\label{fig0330}
\end{figure}

{\bf Superposition state} ${|\textbf{SP}3\rangle}$: Fig. \ref{fig0330} presents the QC with the input of $|\rm{SP}3\rangle$ for various degrees of disorder. Different from the case of $|\rm{SP}2\rangle$ ranging between $-1$ and $0$, $|\rm{SP}3\rangle$ lies between $-1$ and $1$, which elucidates that the scattered modes can be either correlated or anticorrelated. Similarly, it is found that $\bar{C}_{\alpha \beta}$ increases with the increase of $s$.

Compared between $|1,1,1\rangle$ and $|\rm{SP}3\rangle$, it is found that with the same disorder strength, the corresponding distributions of the QC behave differently, although the averages are nearly equal. Therefore, the superpostion affects the distributions of the quantum correlations.

\subsection{The averages and ``deviations" of the quantum correlations}

\begin{figure}[tbhp]
\begin{center}
\includegraphics[width=1.1\textwidth]{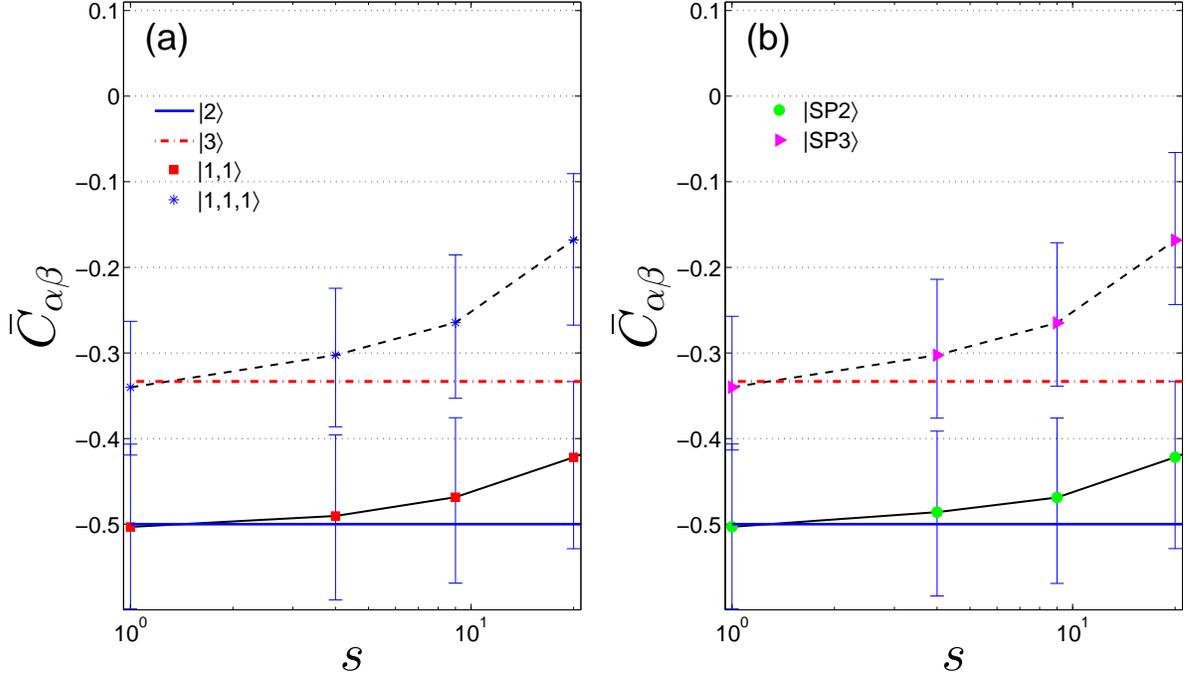} {}
\end{center}
\caption{The averages and ``deviations" of the QC as a function of $s$ for different inputs: (a) product states of $|1,1\rangle$ and $|1,1,1\rangle$; (b) superposition states of $|\rm{SP}2\rangle$ and $|\rm{SP}3\rangle$. 
Each ``deviation" is composed of two parts: one is the positive ``deviation" ($\Delta C_{\alpha \beta}^{P} \equiv \sqrt{\overline{(C_{\alpha \beta} - \bar{C}_{\alpha \beta})^2}}$ if $C_{\alpha \beta} > \bar{C}_{\alpha \beta}$, the part of the ``deviation" above the averaged QC) and the other is the negative ``deviation" ($\Delta C_{\alpha \beta}^{N} \equiv \sqrt{\overline{(C_{\alpha \beta} - \bar{C}_{\alpha \beta})^2}}$ if $C_{\alpha \beta} < \bar{C}_{\alpha \beta}$, the part of the ``deviation" below the averaged QC). The ``deviations" have been minified in a scale of $1:3$ for avoiding overlap with each other}
\label{figcab2s}
\end{figure}

To quantitatively understand the distributions of the QCs with various
state inputs, the averages and ``deviations" of the QCs are monitored in
Fig. \ref{figcab2s}, including the products and superpositions of Fock states.

In Fig. \ref{figcab2s}(a), one can see clearly that the QC trends from $C_{\alpha \beta} < 0 $ to $C_{\alpha \beta} >0$ as the number of input modes increases. Comparing between Figs. \ref{figcab2s}(a) and \ref{figcab2s}(b), one can find that although the superposed state $|\rm{SP}3\rangle$ and the product state $|1,1,1\rangle$ have roughly equal averaged QCs, the distributions of QCs behave differently. For the product state $|1,1,1\rangle$, the negative ``deviation" is larger than the positive one with respect to each $s$, but the behavior is completely opposite for the superposed state $|\rm{SP}3\rangle$ (for detailed numerical results, see table \ref{table001a}). Moreover, we have also examined for the input cases of $|1,1,1,1\rangle$, $|1,1,1,1,1\rangle$, $|\rm{SP}4\rangle$, and $|\rm{SP}5\rangle$, where $|\rm{SP}4\rangle \equiv (|4,0,0,0\rangle + |0,4,0,0\rangle + |0,0,4,0\rangle + |0,0,0,4\rangle)/2$ and $|\rm{SP}5\rangle \equiv (|5,0,0,0,0\rangle + |0,5,0,0,0\rangle + |0,0,5,0,0\rangle + |0,0,0,5,0\rangle + |0,0,0,0,5\rangle)/\sqrt{5}$. Similarly, the superposed states and the corresponding product states have nearly equal averages of the QCs, whereas they present different distributions of the QCs. For clarity, the results are not present in this literature. Besides, for these multi-mode-state inputs, as the disorder is increased, the ensemble-averaged QC increases, indicating that the probability that two photons arrive at two different output ports increases. 

\begin{table*}[tbhp]
\tabcolsep 2mm
\doublerulesep 8mm
\caption{The comparison of positive and negative ``deviations" between the cases of $|1,1,1\rangle$ and $|\rm{SP}3\rangle$ \label{table001a}}
\begin{center}
\renewcommand\arraystretch{2.2}
\begin{tabular}{|c|c|c|c|c|c|c|c|c|}
\hline
Input state& \multicolumn{4}{c|}{$|1,1,1\rangle$} &  \multicolumn{4}{c|}{$|\rm{SP}3\rangle$}\\\cline{1-9} 
$s$ &$1$ &$4$ &$9$ &$20$ &$1$ &$4$ &$9$&$20$ \\ \hline
Positive ``deviation" &  ${0.231}$ & ${0.234}$  & ${0.236}$& ${0.232}$  & ${0.248}$ & ${0.265}$ & ${0.281}$ & ${0.307}$ \\ \hline
Negative ``deviation" &  ${0.237}$ &${0.251}$ & ${0.265}$ & ${0.298}$ &  ${0.220}$ &${0.221}$  & ${0.222}$ & ${0.225}$ \\
\hline
\end{tabular}
\end{center}
\end{table*} 
 
\section{Comparison and discussion}

\subsection{Gaussian state inputs}

Since the cases that we investigate are non-Gaussian-state inputs (products and superpositions of Fock states), we wonder how Gaussian states affect the QCs, where the Gaussian state is defined as a state whose distribution function in phase space (Wigner function) is in the Gaussian form \cite{b2004,wang2007,weedbrook2012,adesso2014}. For instance, the typical Gaussian states include: (1) coherent state, (2) thermal state, and (3) squeezed-vacuum state.

To examine the effects of Gaussian states on the QCs, we investigate the cases of several typical Gaussian states as inputs: (1) coherent state, (2) thermal state, and (3) squeezed-vacuum state.

{\bf Coherent state}: When a coherent state is considered as input, the QC is worked out as zero regardless of the single-mode or multi-mode input state. It means that, with coherent states as input, $C_{\alpha \beta}$ is always a constant of zero \cite{note} as expected. This indicates that output modes are uncorrelated with any coherent state inputs. In other words, any two scattered modes will be simply a product of two coherent states. In addition, it is worthy pointing out that the case of the single-mode state is worked out analytically while the case of the multi-mode state numerically. 

\begin{figure}[tbhp]
\begin{center}
\includegraphics[width=.680\textwidth]{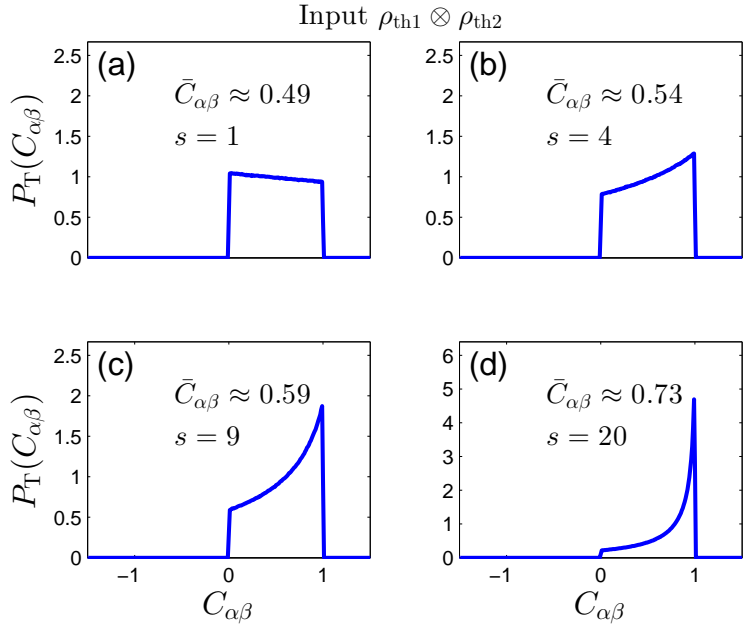} {}
\end{center}
\caption{The normalized probability distribution of the QC with the input of two-mode-thermal state of $\rho_{{{\rm{th}}1}} \otimes \rho_{{{\rm{th}}2}}$ for different disorder strengths: (a) $s = 1$, (b) $s = 4$, (c) $s = 9$, and (d) $s = 20$. The mean photon number of each thermal state is two. Note that y-axis ranges from $0$ to $6$ in (d) while (a)-(c) are from $0$ to $2.5$. Therefore the area under the curve in (d) seems smaller than the ones in (a)-(c). All the other parameters are the same as the case in Fig. \ref{fig11two}}
\label{figth}
\end{figure}

{\bf Thermal state}: If a thermal state is injected, with the single-mode input, one finds the QC being a constant of one analytically. However, when the input is in a multi-mode state, the quantum correlation is not a constant but behaves with a certain probability distribution. In particular, Fig. \ref{figth} plots the distribution pattern of the QC with the two-mode-thermal-state input of $\rho_{{{\rm{th}}1}} \otimes \rho_{{{\rm{th}}2}}$ (the mean photon number of each thermal state is two). It is easily found that the QC lies between zero and one regardless of the degree of disorder. Moreover, with increasing $s$, $C_{\alpha \beta}$ increases and approaches one which reveals that the output modes trend to be strong correlated. 

\begin{figure}[tbhp]
\begin{center}
\includegraphics[width=.680\textwidth]{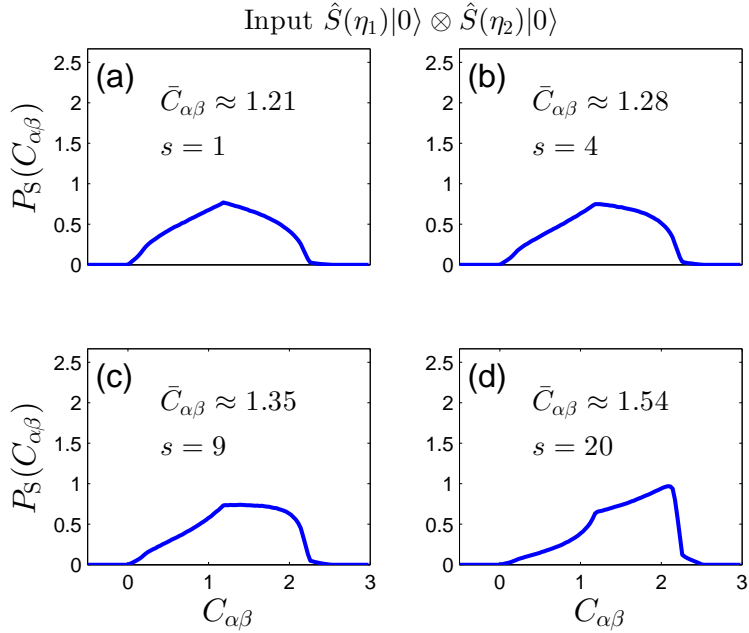} {}
\end{center}
\caption{The normalized probability distribution of the QC with the input of two-mode-squeezed-vacuum state of $\hat{S}(\eta_1) |0\rangle \otimes \hat{S}(\eta_2) |0\rangle$ for different disorder strengths: (a) $s = 1$, (b) $s = 4$, (c) $s = 9$, and (d) $s = 20$, where $\hat{S}(\eta_j) = \exp[(-\eta_j \hat{c}_j^{\dagger 2} + \eta_j^{\ast} \hat{c}_j^2)/2]$ with $\eta_j = r_j \exp(i \theta_j)$ ($j = 1, 2$). $\eta_j^{\ast}$ is the conjugate of $\eta_j$. $\hat{c}_j^{\dagger}$ ($\hat{c}_j$) means the creation (annihilation) operator of the mode $j$. The real parameters, $r_j$ and $\theta_j$ ($j=1, 2$), denote squeezing strength and angle, respectively. The mean photon number of squeezed-vacuum-state is given by $\bar{n}_{{sqz}} = \sinh^2 r_j = 2$ ($ j = 1, 2$). All the other parameters are the same as the case in Fig. \ref{fig11two}}
\label{figsqz}
\end{figure}

{\bf Squeezed-vacuum state}: If the input is in a squeezed-vacuum state, with the single-mode state, the QC is calculated as a constant of $2 + 1 / \bar{n}$ analytically (always large than two) where $\bar{n}$ is the mean photon number of the input beam. Thus, with the increase of $\bar{n}$, the QC decreases. Nevertheless, with the multiple-mode-state input, one has the quantum correlation with a certain distribution pattern. Fig. \ref{figsqz} plots the statistical distribution of the QC with the two-mode-squeezed-vacuum-state input. Note that the QC is always large than zero with various $s$ as shown in Fig. \ref{figsqz}, revealing that the output modes are always correlated. Furthermore, $\bar{C}_{\alpha \beta}$ becomes large with the increase of the disorder, which indicates that large disorder would induce strong correlated output modes.

\subsection{Comparison between Gaussian states and non-Gaussian states}

\begin{table*}[thbp]
\tabcolsep 2mm
\doublerulesep 8mm
\caption{The distribution of the quantum correlation between the scattered modes with the Gaussian and non-Gaussian state inputs in disordered media. The second-order quantum correlation of single-mode input state is defined as $C = \langle \hat{a}^{\dagger} \hat{a}^{\dagger} \hat{a} \hat{a} \rangle /(\langle \hat{a}^{\dagger} \hat{a}\rangle \langle \hat{a}^{\dagger} \hat{a}\rangle) - 1$ \cite{note} with $\hat{a}^{\dagger}$ ($\hat{a}$) being the creation (annihilation) operator of the input mode. $P_{i}(C_{\alpha \beta})$ (i = \{T, S, F, and SF\}) represents the distribution of the QC with respect to the each input. $\bar{n}$ denotes the mean photon number of the corresponding input state. The ``NG'' stands for not given due to the multiple modes of the superposition of Fock state \label{Tab002}}
\begin{center}
\renewcommand\arraystretch{2.2}
\begin{tabular}{cccccc}
\hline
\multicolumn{1}{|c|}{Input state} &\multicolumn{3}{c}{Gaussian} & \multicolumn{2}{c|}{Non-Gaussian}  \\ \cline{2-6}

\multicolumn{1}{|c|}{}& \multicolumn{1}{c}{Coherent} & \multicolumn{1}{c}{Thermal} & \multicolumn{1}{c|}{\renewcommand\arraystretch{1.1}\begin{tabular}{@{}c@{}}Squeezed \\ vacuum\end{tabular}} & \multicolumn{1}{c}{\renewcommand\arraystretch{1.1}\begin{tabular}{@{}c@{}}Product \\ of Fock\end{tabular}} & \multicolumn{1}{c|}{\renewcommand\arraystretch{1.1}\begin{tabular}{@{}c@{}}Superposition \\ of Fock\end{tabular}}   \\\cline{1-6}

\multicolumn{1}{|c|}{Single-mode} & \multicolumn{1}{c|}{$0$ \cite{starshynov2016}} & \multicolumn{1}{c|}{$1$ \cite{starshynov2016}} & \multicolumn{1}{c|}{$2+\frac{1}{\bar{n}}$ \cite{starshynov2016}} & \multicolumn{1}{c|}{$-\frac{1}{\bar{n}}$ \cite{ott2010}} & \multicolumn{1}{c|}{NG}\\ \cline{2-6}

\multicolumn{1}{|c|}{Multi-mode} & \multicolumn{1}{c|}{$0$ \cite{starshynov2016}} & \multicolumn{1}{c|}{$P_{\rm{T}}(C_{\alpha \beta})$} & \multicolumn{1}{c|}{$P_{\rm{S}}(C_{\alpha \beta})$} & \multicolumn{1}{c|}{$P_{\rm{F}}(C_{\alpha \beta})$} &  \multicolumn{1}{c|}{$P_{\rm{SF}}(C_{\alpha \beta})$}  \\ \hline

\multicolumn{1}{|c|}{\renewcommand\arraystretch{1.1}\begin{tabular}{@{}c@{}}Second-order \\ quantum correlation\end{tabular}} & \multicolumn{1}{c|}{$0$ \cite{walls2007}} & \multicolumn{1}{c|}{$1$ \cite{walls2007}} & \multicolumn{1}{c|}{$2+\frac{1}{\bar{n}}$ \cite{barnett2002}} & \multicolumn{1}{c|}{$-\frac{1}{\bar{n}}$  \cite{barnett2002}} & \multicolumn{1}{c|}{NG} \\ \hline

\end{tabular}\\

\bigskip
{}
\end{center}

\end{table*}

In summary, table \ref{Tab002} presents the comparison between the Gaussian and non-Gaussian states where there are three kinds of Gaussian states (coherent, thermal, and squeezed-vacuum states) and two types of non-Gaussian states (products or superpositions of Fock states) considered. Clearly, each single-mode-state input has a constant QC regardless of the input being Gaussian or non-Gaussian state. Particularly, for the squeezed-vacuum or product of Fock states, the constant QC depends only on the incident mean total photon number, which is not true for the cases of coherent and thermal states. In addition, it is interesting that the constant QC between output modes is equivalent to the second-order quantum correlation of the corresponding single-mode-input state as depicted in table \ref{Tab002}, where the corresponding proof is presented in Append. \ref{qcappd}.

In contrast, the multi-mode-state inputs have the QC with a certain probability distribution except for the coherent state. As a matter of fact, for the multi-mode-coherent-state input, the scattered modes are uncorrelated ($C_{\alpha \beta} = 0$). In addition, the Gaussian states always present the non-negative QC while the non-Gaussian states show the QC which can be either positive or negative.

\section{Conclusion}

In conclusion, we investigate the distribution of the quantum correlation between the scattered modes in the different-disordered media with various input states (Gaussian and non-Gaussian states). It is revealed that for the initial quantized light, both internal degrees (e.g. photon number distribution on the Fock basis) and external degrees (e.g. number of spatial modes) play significant roles in the quantum correlations between the scattered modes.

It is discovered that with the single-mode-state input, regardless of the Gaussian or non-Gaussian states, the quantum correlation between output modes is always a constant which is equivalent to the value of the second-order quantum correlation of the input beam. In particular, the single-mode-squeezed-vacuum (Fock-state) input has a constant quantum correlation, $2+1/\bar{n}$ ($-1/\bar{n}$), which is related to the incident mean total photon number $\bar{n}$. In contrast, the single-mode-coherent (thermal) state input has a constant quantum correlation, zero (one), which does not depend on the input photon number. Meanwhile, if the input is in a multi-mode state, the coherent state still has a constant quantum correlation of zero whereas the other states show the quantum correlation with a certain probability distribution. As the disorder is increased, the averaged quantum correlation increases for most situations except for single-mode-state and multi-mode-coherent-state inputs. These results may find applications in quantum information processing, such as quantum-secure authentication \cite{goorden2014}.

\noindent {\bf Acknowledgment} We thank Prof. Jun-Hong An for helpful discussions and Prof. Song Sun for insightful suggestions. This work was supported by the Science Challenge Project (Grant No. TZ2018003-3) and National Natural Science Foundation of China (Grant Nos. 61875178 and 11605166).

\appendix
\section{Method for generating the random scattering matrix}\label{rmtappd}

\begin{figure}[thbp]
\begin{center}
\includegraphics[width=.90\textwidth]{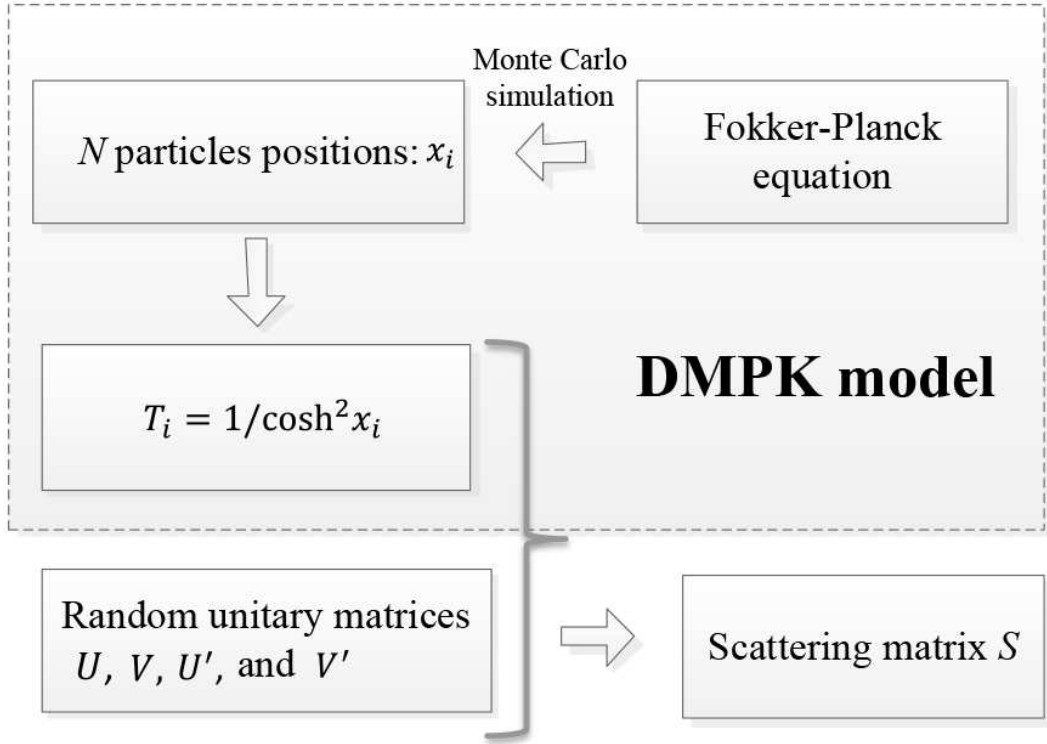} {}
\end{center}
\caption{The outline of calculation of random scattering matrix $S$ including three steps: (I) Calculating $\cal{T}$ by the Monte Carlo simulation solving the DMPK model; (II) Generating random unitary matrices $U$ ($U'$) and $V$ ($V'$); (III) Determining $S$ through Eq. (\ref{sm})}
\label{dmpk}
\end{figure}

This section presents how to generate the random scattering matrix $S$. According to Ref. \cite{beenakker1997}, the scattering matrix $S$ can be decomposed as
\begin{align}
S  =\left(
\begin{array}
[c]{cc}%
U & 0\\
0 & V%
\end{array}
\right)
\left(
\begin{array}
[c]{cc}%
-\sqrt{1-\cal{T}} & \sqrt{\cal{T}} \\
\sqrt{\cal{T}} & \sqrt{1-\cal{T}}%
\end{array}
\right)
\left(
\begin{array}
[c]{cc}%
U' & 0\\
0 & V'%
\end{array}
\right),\label{sm}
\end{align}
where $U$ ($U'$) and $V$ ($V'$) are random unitary matrices and $\cal{T}$$\equiv$ $\text{diag}(T_1, T_2,..., T_N)$ is the set of transmission eigenvalues ($N$ is the number of transmission channels). As shown in Fig. \ref{dmpk}, the scattering matrix $S$ can be obtained by the three steps: (I) Calculating $\cal{T}$ by the approach of Dorokhov, Mello, Pereyra, and Kumar (DMPK) \cite{froufe-perez2002,froufe-perez2005}; (II) Generating random unitary matrices $U$ ($U'$) and $V$ ($V'$); (III) Determining $S$ through Eq. (\ref{sm}).

 In the step (I), a Monte Carlo solution of the DMPK model \cite{froufe-perez2002} is utilized to calculate $\cal{T}$. The DMPK approach describes the evolution of joint probability distribution of transmission eigenvalues $P(\{T_i\})$ in the form of a $N$-dimensional Fokker-Planck equation \cite{froufe-perez2002}. According to Ref. \cite{froufe-perez2002}, in terms of the variables ($\{x_i\}$), the transmission eigenvalue can be written as $T_i\equiv 1/\cosh^2 x_i$. And the results can be expressed in the form of a Gibbs distribution
\begin{align}
P(\{x_i\}) \propto \exp [ -\beta H(\{ x_i\})],
\end{align}
where
\begin{align}
H(\{ x_i\}) &= \sum_{i<j} U(x_i,x_j) + \sum_{j} V(x_j),\\
U(x_i,x_j) &= -\frac{1}{2} (\ln |\sinh^2 x_i - \sinh^2 x_j| + \ln|x_i^2 - x_j^2|),\\
V(x) &= \frac{\gamma}{2 \beta s} x^2 - \frac{1}{2 \beta} \ln|x\sinh 2x|,
\end{align}
with $\gamma = \beta (N-1) + 2$ and $s= L/l$ ($l$ being the mean free path and $L$ the length of disordered medium). $H(\{ x_i\})$ may be regarded as the Hamiltonian function of $N$ particles at positions $x_i$ in one dimensional space, $U(x_i,x_j)$ the interaction potential and $V(x)$ the effective confining potential \cite{froufe-perez2002}. In particular, we concentrate on the case of time reversal, namely $\beta = 1$, $U' = U^{T},$ and $V' = V^{T}$ ($U^{T}$ and $V^{T}$ denoting the transpose of $U$ and $V$, respectively) \cite{beenakker1997}.

To evaluate particle positions ($\{x_i\}$), the Monte Carlo simulation is performed as illustrated below \cite{canali1996}. One partitions the one dimensional real axis into bins with boundaries $x_n$ given by
\begin{align}
x_n = n {\rm{\Delta}} x,\quad\quad n = 1,...,N,
\end{align}
where $\rm{\Delta}$$x$ is the width of the bins. Initially, each particle is located at the center of each bin. In the system, the positions ($\{x_i\}$) are evolved under the Hamiltonian $H$. For the evolution of particle positions, a simple Metropolis algorithm is performed. At each time step, we monitor the all particles and attempt to move each one. In fact, in each time step, we pick $N$ times one particle at random among the all particles, thus it is possible that one particle is touched more than once and another is not chosen in one particular sweep. The moving attempt can be described as: (a) picking at random any position between the particle that proceeds and the one that follows the particle that we are trying to move; (b) and taking this position as the new attempted position. We choose the attempted move in this particular way simply to optimize the convergence rate of the algorithm. An important property is that if it starts with an order sequence of particles, $x_1 < x_2 < ...< x_N$, the sequence remains ordered in each time step. 

Whether or not to accept the move is dependent on the change ${\rm{\Delta}} E$ to the system energy that would occur if the particle were moved to the new position. If ${\rm{\Delta}} E$ is negative, the move is accepted, and let the particle be in the new position. If ${\rm{\Delta}} E$ is positive, the move is accepted conditionally. A random number between zero and one is generated and the move would be accepted if this number is smaller than $\exp(-\beta {\rm{\Delta}} E)$. Before recording any quantity (positions $\{x_i\}$), the system should reach equilibrium which can be obtained by a certain bunch of warming-up sweeps.

The simulations are performed over systems with up to $20$ particles. The equilibration is usually reached very fast and $10^5$ sweeps is used to warm up the system. The values of $(\{x_i\})$ are recorded over $10^7$ sweeps and the statistics that we are able to obtain are usually excellent. After ($\{x_i\}$) is worked out, one can easily arrive at $\cal{T}$ by evolving $\cal{T}=$$\text{diag}(T_1, T_2,..., T_N)$ with $T_i\equiv 1/\cosh^2 x_i$. 

In step (II), the random unitary matrices $U$ $(U^T)$ and $V$ $(V^T)$ are generated, which is followed by the step (III), according to Eq. (\ref{sm}), one then obtain the scattering matrix $S$.

\section{Proof}\label{qcappd}
We proof that, in the case of single-mode-state input, the two-channel quantum correlation $C_{\alpha \beta}$ between output modes is equal to the second-order quantum correlation of the single-mode input beam where the  second-order quantum correlation is defined by
\begin{align}
C = \dfrac{\langle \hat{a}_i^{\dagger} \hat{a}_i^{\dagger} \hat{a}_i \hat{a}_i \rangle} {\langle \hat{a}_i^{\dagger} \hat{a}_i\rangle \langle \hat{a}_i^{\dagger} \hat{a}_i\rangle} - 1,
\label{csecondorder}
\end{align}
where $\hat{a}_i^{\dagger}$ ($\hat{a}_i$) denotes the creation (annihilation) operator of the incident mode $i$. Note that, there is a small difference where the second-order QC introduced here has an additional minus one compared with the traditional definition \cite{walls2007}. The corresponding QC between scattered modes is found to be
\begin{align}
C_{\alpha \beta} &= \frac{\sum_{mnpq} S_{\alpha m}^{\ast} S_{\beta p}^{\ast} S_{\beta q} S_{\alpha n} \langle \hat{a}_m^{\dagger} \hat{a}_{p}^{\dagger} \hat{a}_{q} \hat{a}_n \rangle}{\left(\sum_{mn} S_{\alpha m}^{\ast} S_{\alpha n} \langle \hat{a}^{\dagger}_{n} \hat{a}_{m} \rangle\right) \left( \sum_{pq} S_{\beta p}^{\ast} S_{\beta q} \langle \hat{a}^{\dagger}_{p} \hat{a}_{q} \rangle \right)}-1.
\label{cab1mode}
\end{align}
Due to the single-mode-state input, Eq. (\ref{cab1mode}) can be then reduced to
\begin{align}
C_{\alpha \beta} &= \frac{ S_{\alpha i}^{\ast} S_{\beta i}^{\ast} S_{\beta i} S_{\alpha i} \langle \hat{a}_i^{\dagger} \hat{a}_{i}^{\dagger} \hat{a}_{i} \hat{a}_i \rangle}{\left( S_{\alpha i}^{\ast} S_{\alpha i} \langle \hat{a}^{\dagger}_{i} \hat{a}_{i} \rangle\right) \left(  S_{\beta i}^{\ast} S_{\beta i} \langle \hat{a}^{\dagger}_{i} \hat{a}_{i} \rangle \right)}-1\\ \nonumber
&=\dfrac{\langle \hat{a}^{\dagger}_i \hat{a}^{\dagger}_i \hat{a}_i \hat{a}_i \rangle} {\langle \hat{a}^{\dagger}_i \hat{a}_i\rangle \langle \hat{a}^{\dagger}_i \hat{a}_i\rangle} - 1,
\end{align}
where it is the same as the result in Eq. (\ref{csecondorder}).


\begin{thebibliography}{99}
\bibitem{noteinput} Note: for the sake of simplicity, single-mode input $|n\rangle$ stands for $|n\rangle \otimes (|0\rangle^{\otimes 2N-1})$ which denotes that one input mode is in Fock state $|n\rangle$ while the other $N-1$ input modes are in vacuum states $|0\rangle$. Similarly, in the following context, two-mode (three-mode) input state indicates that the other $2N-2$ ($2N-3$) input modes are in vacuum states $|0\rangle$. In fact, a disordered waveguide of $N$ transmission channel has $2N$ input modes, where $N$ modes are from the left-hand side and the other $N$ modes from the right-hand side.
\end{thebibliography}
\end{document}